# Has Metallic Hydrogen Been Made in a Diamond Anvil Cell?

W.J. Nellis[1], Arthur L. Ruoff[2], and Isaac F. Silvera[1]
[1]Lyman Laboratory of Physics, Harvard University, Cambridge, MA 02138
and
[2]Department of Materials Science and Engineering, Cornell University, Ithaca, NY 14853

Eremets and Troyan recently reported a transition of molecular hydrogen to metallic hydrogen (MH) at a pressure of ~270 GPa and 300 K. The quest for MH has been going on since 1935 when Wigner and Huntington predicted solid $H_2$ at 0 K would transition to solid H with an associated insulator-metal transition at high pressure. Since then, observation of MH in solid hydrogen has yet to be confirmed. Because of its scientific importance we have considered their experimental evidence to determine if their claim is justified. Based on our analysis there is no evidence for MH in their experiments.

The purpose of this paper is to answer the question posed in the title. In a recent Letter by M. I. Eremets and I. A. Troyan (ET) [1] and in an associated News and Views by Jephcoat [2], metallic hydrogen (MH) is reported to have been made in a diamond anvil cell (DAC-see Fig. 1) at a pressure of 260-270 GPa and 300 K. ET's article reports experimental results in a challenging regime for study of matter under extreme conditions and presents optical data showing new phase transitions that appear if the sample is studied at room temperature, in contrast to previous studies at lower temperatures. Observation of MH is based principally on measurements of resistance, but also on Raman scattering, and sample photographs in reflection and transmission.

The quest for metallic hydrogen has been going on for over one hundred years. Prior to the production of the condensed phases many scientists expected that condensed hydrogen would be a metal, similar to alkali metals below hydrogen in the first column of the Periodic Table. Then hydrogen was liquefied in 1898 and solidified in 1899 by James Dewar [3]; both phases are transparent insulators. We now know that hydrogen is a homonuclear diatomic molecular solid below 13.9 K in equilibrium with its vapor pressure [4].

After Quantum Mechanics had been established, Wigner and Huntington predicted in 1935 that solid molecular $H_2$ at 0 K would dissociate to an atomic solid H at a pressure estimated to be about 25 GPa [5]. This structural transition would be accompanied by an insulator-metal transition (IMT). The high-pressure phase has been predicted to be a possible high temperature superconductor (SC) [6, 7]. Calculations of the transition pressure for the metallic phase have challenged theorists, but modern analysis, reviewed elsewhere [8], now considers that the dissociative transition will take place in the 400 to 600 GPa region. At lower pressures in the region 200-400 GPa, molecular hydrogen consisting of paired protons has been predicted to become a conductor, and also a high temperature SC [9-11]. Calculations predict that metallic hydrogen may be metastable with a potential barrier that prevents it from returning to the molecular phase when the pressure is released [12]. Exciting applications exist if this is so, such as room temperature SC materials and powerful rocket propellants [13]. Theory also predicts that due to the large zero-point energy, hydrogen may be an atomic liquid at T=0 K and very high pressure [14]. Thus, the predicted trajectory with increasing pressure is that hydrogen becomes a solid molecular metal, then dissociates to an atomic metal that may also be in the liquid state and may remain MH when pressure is released. For these reasons there have been ongoing experimental efforts for decades to produce MH under challenging conditions. In their analysis, ET do not consider the possibility that a metal containing hydrogen pairs may have formed.

Metallic fluid hydrogen has been observed at 140 GPa and 3000 K via a Mott-like transition under multiple-shock compression [15, 16]. Hydrogen has previously been studied in a DAC at room temperature as high as 342 GPa, pressures higher than those reported by ET, without observing MH and showing that it is partially transparent with a band gap near 2 eV [17, 18]. Blackening of hydrogen observed at a lower pressure, 320 GPa [19], has been explained as an issue of pressure calibration [20], although blackening has also been reported at lower pressures [21]. A number of reports of creation of metallic hydrogen under static conditions have been published [22-25], but have since been determined to be incorrect [26-28].

Because of the scientific importance of this issue, we have considered the experimental evidence presented by ET to determine if their claim is justified. Based simply on information provided by them [1], we find no direct evidence to indicate that MH was produced in their DAC.

We first consider their claim for MH, which is based on measurements of electrical resistance R. In fact, neither measured electrical resistivity nor frequency dependent reflectivity, both key physical properties, are given in their paper.

The first thing to consider is whether ET's measured resistances are in fact expected for MH. Fig. 3a of ET shows a sharp drop in resistance R from $10^8$ Ω to ~$10^7$ Ω near 235 GPa and at 270 GPa, a larger drop from $10^7$ to $10^4$ Ω. ET interpret the latter as a transition to metallization with a resistance of $10^4$ Ω (and ignore the first abrupt drop in resistance). However, neither the value nor drop of resistance R carry information about metallization. It is essential to compare electrical resistivity ρ, a material property, as R can take on any value depending on sample geometry. The defining equation of ρ for a rectangular sheet conductor is R=ρ(L/A), where L/A is a geometrical factor, L is the length, and A is the cross sectional area of current flow parallel to L. In the geometry of ET's DAC experiments current flow is not perfectly parallel to L but corrections to the geometric factor are of order a factor of 2, which is essentially negligible because we are concerned with many orders of magnitude in the resistivity of hydrogen, rather than precise values.

In Fig. S6 ET give dimensions for their claimed MH sample, with R=$10^5$ Ω. From the figure caption, L=2 μm and A=80 μm$^2$; thus, ρ=4x$10^8$ μΩ-cm. For a good metal such as Cu at 300 K, ρ=1 μΩ-cm, a factor of 4x$10^8$ smaller than implied by the measured R at ~262 GPa and room temperature. Fig. 3 shows a larger drop in resistance for a different sample; if we apply the same geometric factors we find ρ=4x$10^7$ μΩ-cm. If ET's MH were a poor-metallic liquid with a resistivity of 500 μΩ-cm [16], then their measured resistance would be a factor of $10^6$ larger than expected for this geometry (ignoring contact resistance, which we take up in the following paragraph). Thus, if ET indeed had MH in their electrical circuit in their cell, their measured values of R would be several orders of magnitude smaller than observed. The claim that MH was observed is not validated by their measured values of R.

In the caption to Fig. 3, ET estimate that contact resistance might be ~300 Ω, which would be in series with the hydrogen sample. Since putative resistance of a MH sample is negligible compared to 300 Ω (see below), the expected "sample" resistance would also be ~300 Ω. The measured resistance of $10^4$ ohms would then be ~30 times greater than expected and 9700 Ω are unaccounted for. Thus, it is unlikely that metallic hydrogen is the cause of the measured changes in R, even if a huge contact resistance is included in series with an MH sample.

In the above two cases, current was assumed to flow only through the hydrogen sample. However, the hydrogen sample and gasket are connected as parallel resistances. In the case considered here, we assume that hydrogen remains a perfect insulator and leakage current flows only through the epoxy-alumina gasket. In this case, leakage current would flow a distance of L=2 μm, but over a much larger area of the gasket surfaces than the ~80 μm$^2$ of the hydrogen sample. Thus, current flow through the gasket thickness with metal current leads on opposite faces is a possible explanation for measured R=$10^4$ Ω, without the presence of MH.

ET therefore need to demonstrate experimentally that resistance of their basic design and gasket, without the gasket hole and dense hydrogen, is large compared to resistance measured with the gasket hole and hydrogen, a challenging requirement if diffusion of hydrogen is involved. They have studied the $Al_2O_3$-epoxy gasket in the absence of hydrogen at room temperature and 300 GPa, but do not give values of its resistance. Dense hydrogen is expected to diffuse into the epoxy [29]; their evidence against this, a lack of a hydrogen Raman signal from the epoxy is expected, as the hydrogen concentration in the gasket would be low. Moreover, hydrogen may have chemically reacted with the epoxy and not be in the molecular form. They have not shown the gasket to be insulating in the presence of hydrogen.

Another question is whether their electrical circuit was sufficiently sensitive to be able to observe metallic hydrogen at all? If their hydrogen sample was a good metal with $\rho=1$ $\mu\Omega$-cm or a very poor metal with $\rho=10^3$ $\mu\Omega$-cm, then its resistance would be 0.25 m$\Omega$ or 0.25 $\Omega$, respectively. Since their measured resistances of putative MH are $10^4$ $\Omega$ and $10^5$ $\Omega$ from Fig. 3a and S6b, respectively, it is doubtful that the electrical diagnostic system could resolve a sub-$\Omega$ resistance out of $10^4$ $\Omega$ to $10^5$ $\Omega$. The same is true with a contact resistance of 300 $\Omega$. That is, based on their paper ET's diagnostic system does not have the sensitivity to detect the presence of metallic hydrogen or anything else in the gasket hole.

As ET point out, a rigorous proof that a material is a metal is to show that the electrical conductivity remains finite in the limit that T goes to zero [30]. While the discussion above shows R is virtually insensitive to the presence of hydrogen in the gasket hole and their temperature dependences of R in Figs. 3 and S6, do not show metallic behavior, ET use their R-T data to analyze hydrogen. In Fig. S6 (caption) they calculate what they call an activation energy $E_i$, and find $E_i = 0.3$ meV from the temperature dependence of R at 275 GPa. Metals, such as MH, do not have activation energies, but semiconductors do. In this case they should fit their data to $\ln R / R_0 = E_{gap} / 2kT$ (rather than $\ln R / R_0 = E_i / kT$), which yields $E_{gap} / k = 1.75$ K. To demonstrate metallic behavior their thermal studies would need to probe the region of T~1-2 K, whereas measurements stopped at 30 K. (We note that for the sample in Fig. 3 $E_i = 8$ meV, rather than 0.3 meV). We conclude that the temperature dependence of the resistance is not consistent with a metal, but has the behavior of a narrow gap semiconductor.

It is particularly disturbing that ET report conductance in only 5 of their 70 runs. A claim to have made MH should explain why in 93% of their runs, conductivity is not observed. Their claimed observation of metallic hydrogen in the other 7% of experiments might well be the result of spurious, not-understood causes. It would be important to show resistance vs. pressure on loading and unloading from ~300 GPa for the two runs; for example, resistances shown in Figs. 3 and S6, to show reproducibility or hysteresis. We note that the Raman spectra of Figs. 1 and 2 indicate a phase transition at about 220 GPa, while Fig. S6 indicates (the same) phase transition at 239 GPa, the pressure at which a resistance drop occurs, shown in Fig 3. Another observation in Fig. S6d is that after laser illumination the resistance does not recover. The laser may have promoted hydride formation of the electrodes or dissolution into the hydrogen, discussed next.

One can speculate as to alternate explanations for the observations. Implicit in their discussions is that hydrogen above 200 GPa pressure does not react chemically with any of its surrounding materials. However, hydrogen is an extremely reactive material and so room temperature hydrogen might well react above ~200 GPa, which might be the cause of some of their observations. For example, experiments have reported the dissolution of metals into hydrogen at modest to high pressures and temperatures [31, 32], which could lead to conductivity in the hydrogen [33]. Is it not likely that electrical leads made of copper [34], gold [35], or tantalum [36] (all of which have hydrides at high pressure) become hydrides and also partially dissolve (reversibly) into the hydrogen to conduct and reflect?

Diamonds are hard but still deformable. It is well-known that under pressure, the diamond culet flat cups so that the edges of the flats of opposing diamonds are closer than the centers (see Fig. 1). Observation [37] as well as finite element analysis [38] demonstrates this for beveled diamonds. Scaling the results of Ref. [38] to the conditions of ET yields edges touching when center thickness is 1.7 microns; a slight misalignment of diamond anvils promotes this channel for shorting of the leads. It is possible that the leads touch to conduct, and separate when pressure is reduced. A partial test for this would require a gasket with a hole and a compressible sample such as helium, which is not expected to metallize, replacing the hydrogen. The procedure described in the caption of Fig. S6 would not test for this problem as both leads are on one anvil.

ET point out that their measurements differ from earlier measurements that were carried out at lower temperatures in this pressure region, with the implication that at room temperature the hydrogen in their study can enter a metastable forbidden phase. However, earlier experiments at room temperature in this same pressure region using tungsten gaskets (which were shown not to dissolve into hydrogen at that temperature [18]) showed no sign of a metallic, darkening, or reflecting behavior [17]. Moreover, this work included Raman studies at 300 K in the same pressure regime. ET ignored this well-known study that does not support their claim of MH.

ET speculate that their sample is liquid atomic hydrogen. There are experimental methods to establish the liquid phase; no acceptable evidence at all is presented to show that their sample is in the liquid phase. Raman measurements indicate a phase transition in the 220 to 239 GPa region, within the molecular (paired) phase (the fundamental intermolecular vibration persists above this pressure). It is imputed that the disappearance of the molecular Raman spectra above 260 GPa may be evidence that the new phase is atomic or liquid. But, due to the blackening (or reflecting) of the sample at this pressure, the laser beam and the resultant Raman intensities would be highly attenuated resulting in low signal-to-noise. There is no evidence of liquid or dense atomic hydrogen.

Finally, ET show photographs of the sample/leads/gasket as evidence of metallization. That the sample appears black to white light is only evidence that an energy gap has narrowed down into the infrared so that higher energy photons are absorbed due to electronic excitation across the gap. If the imaginary part of the dielectric constant of the sample is large, then the sample will be reflective in the visible, just as Ge and Si look metallic at wavelengths in the visible. To show metallic behavior one must measure over a broad frequency range to establish Drude free-electron behavior [39]. An additional comment relative to the photographs is the following. The typical electrode thickness of 100 nm mentioned in Methods of ET, is inconsistent with their experimental observations. A calculation of transmission and reflection under optimum conditions (hydrogen index matched to diamond at 130 GPa [40]) shows transmittance less than $10^{-3}$ in the visible and reflectance of about 0.9 for the light path through the DAC and sample (air-diamond-Cu-$H_2$-Cu-Diamond-air) at 500 nm (a similar result applies for Au or Ta electrodes). Since photos show the electrodes to be partially transparent, either the electrodes were substantially thinner than reported, were chemically attacked to form hydrides, or partially dissolved into the hydrogen.

In conclusion it is our opinion that the measurements of Eremets and Troyan provide no experimental evidence that metallic hydrogen has been produced in their DACs under conditions of static pressure. In particular measured resistances differ by 6 orders of magnitude or more from those expected for typical metallic resistivities (1 to $10^3$ $\mu\Omega$-cm), while their electrical circuit lacks the sensitivity to resolve expected sub-$\Omega$ resistances of MH. Resistances are not reproducible. The reported temperature dependences support at best semiconductive behavior of their circuit components. No evidence is presented for the speculated atomic liquid phase of the sample. Raman spectra at best support observation of phase transitions in solid molecular hydrogen at room temperature, but this is not a test for an IMT. Photographic evidence of reflectance and transmittance is neither quantitative nor convincing. In fact ET do not publish one measured physical property of metallic hydrogen to substantiate their claim that can be reproduced independently. At this point it is not clear how to modify the phase diagram; we point out that the phase diagram of hydrogen presented by ET in their Fig. 4 is incorrect as it shows phases I, II, and III for deuterium, not hydrogen.

Equal contributions were made by the authors who have extensive experience in the study of hydrogen under pressure. Support from NSF grant DMR-0804378, DoE grant DE-FG52-10NA29656, and NASA NIAC fund 133540 is gratefully acknowledged. We acknowledge assistance from Raymond Greene.

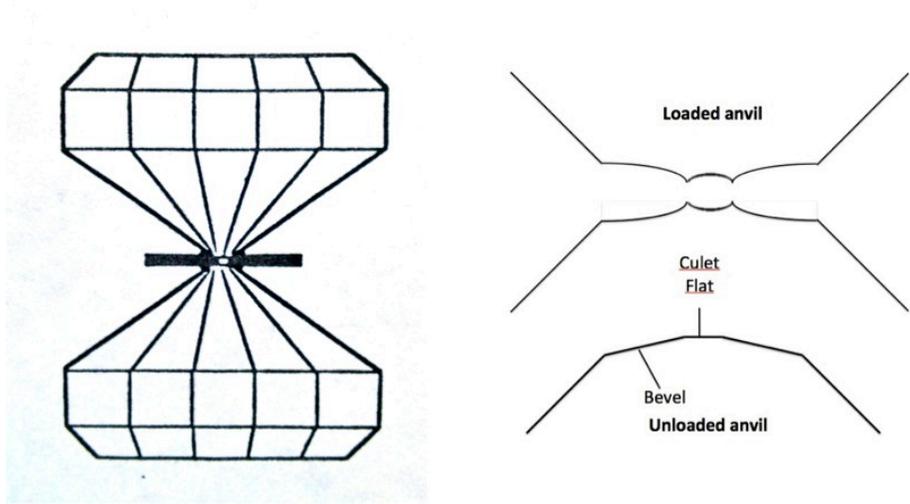

Fig. 1. The heart of a DAC. Two opposing diamond anvils have flats on the tapered culet end that compress the sample confined in the gasket hole. Diamond dimensions are typically a few millimeters. Electrical leads can be deposited on the culet and insulated from the metallic gasket for resistance measurements. Diamond flats are generally surrounded by a bevel, angled at several degrees. On the lower right is a central cross section of an unloaded anvil showing the central flat and a bevel angle of several degrees. Under load, pressing on a gasket, diamonds elastically deform, or cup, as shown.